\documentstyle[aasms4,psfig]{article}

\newcommand\beq{\begin{equation}} 
\newcommand\eeq{\end{equation}}

\received{} 
\accepted{} 

\lefthead{Ma et al.} 
\righthead{Spectral Energy Distributions and Age Estimates of 78 Star Clusters in M33} 

\begin{document}

{\title{Spectral Energy Distributions and Age Estimates of 78 Star Clusters in M33}

\author{ 
Jun Ma\altaffilmark{1}, 
Xu Zhou\altaffilmark{1},
Jiansheng Chen\altaffilmark{1},
Hong Wu\altaffilmark{1},
Zhaoji Jiang\altaffilmark{1}, 
Suijian Xue\altaffilmark{1},
and
Jin Zhu\altaffilmark{1},
}

\altaffiltext{1}{National Astronomical Observatories, 
Chinese Academy of Sciences, Beijing, 100012, P. R. China;
majun@vega.bac.pku.edu.cn}

\authoremail{majun@vega.bac.pku.edu.cn}

\begin{abstract}

In this third paper of our series, we present CCD spectrophotometry of 78 star
clusters that were detected by Chandar, Bianchi, \&
Ford in the nearby spiral galaxy M33.
CCD images of M33 were obtained as a part
of the BATC Color Survey of the sky
in 13 intermediate-band filters from 3800 to 10000{\AA}.
By aperture photometry, we obtain the
spectral energy distributions of these 78 star clusters.
As Chandar, Bianchi, \& Ford did, we estimate the ages of our sample
clusters by comparing the photometry of each object with theoretical
stellar population synthesis models for different  values of
metallicity. We find that the sample clusters formed continuously in
M33 from $\sim 3\times10^6$ -- $10^{10}$ years.
This conclusion is consistent with Chandar, Bianchi, \&
Ford. The results also show that, there are two peaks
in cluster formation, at
$\sim 8\times10^6$ and $\sim 10^9$ years in these clusters.
\end{abstract}


\section{INTRODUCTION}

The importance of the study of star clusters is difficult to
overstate, especially in Local Group galaxies. Star
clusters, which represent, in distinct and luminous ``packets'',
single age and single abundance points, and encapsulate at least
a partial history of the parent galaxy's evolution,
can provide a unique laboratory for studying.
For example, globular clusters can be utilized
to provide a lower limit to the age of the parent galaxy
provided their ages can be ascertained, and to study the
properties of the parent galaxy soon after its formation.

M33 is a small Scd Local Group galaxy, about 15 times farther from
us than the LMC (distance modulus is 24.64) (Freedman,
Wilson, \& Madore 1991; Chandar, Bianchi, \&
Ford 1999a). It is interesting and important because
it represents a morphological type intermediate
between the largest ``early-type'' spirals and the dwarf
irregulars in the Local Group (Chandar, Bianchi, \& Ford 1999a). 
Besides, At a distace of $\sim 840$ kpc, M33 is the only
nearby late-tye spiral galaxy, it can provide an
important link between the cluster populations of earlier-tye
spirals (Milky Way galaxy and M31) and the numerous,
nearby later-type dwarf galaxies.
A database of star clusters for M33 have been
yielded from the ground-based work (Hiltner 1960;
Kron \& Mayall 1960; Christian \& Schommer 1982, 1988;
Melnick \& D'Odorico 1978; \cite{Mochejska98}),
and from the {\it {Hubble Space Telescope
(HST)}} images (Chandar, Bianchi, \&
Ford 1999a; \cite{Chandar01}). Especially,
the $HST$ spatial resolution allowed Chandar, Bianchi, \& Ford (1999a,
2001) to penetrate the crowded, spiral regions of
M33, yielding the unbiased, representative sample of star clusters,
which can be used to probe the global properties of M33.
Since clusters at the distance of M33 are easily distinguished
from stellar sources in $HST$ WFPC2 images, the clusters detected by
$HST$ WFPC2 images are reliable. 

Using the $Hubble$ $Space$ $Telescope$ WFPC2 multiband images
of 20 fields in M33, Chandar, Bianchi,
\& Ford (1999a) detected 60 star clusters in this spiral galaxy.
These clusters
sample a variety of environments from outer regions to spiral
arms and central regions, and are the first unbiased, representive
sample of star clusters in M33. Then, Chandar, Bianchi, \& Ford (1999b)
estimated the ages and masses for these star clusters by comparing the integrated
photometric measurements with evolutionary models and theoretical
$M/L_V$ ratios. They found the 60 star clusters to form continuously
in their parent galaxy from $\sim 4\times 10^6 - 10^{10}$ years, and to have masses
between $\sim 4\times 10^2$ and $3\times 10^5$ $M_{\odot}$.

M33 was observed as part of galaxy calibration program of
the Beijing-Arizona-Taiwan-Connecticut (BATC) Multicolor Sky
Survey (Fan et al. 1996; Zheng et al. 1999)
from September 23, 1995. This program
uses the 60/90 cm
Schmidt telescope at the Xinglong Station of Beijing Astronomical
Observatory (BAO),
and has custom designed a set of
15 intermediate-band filters to do spectrophotometry for
preselected 1 deg$^{2}$ regions of the northern sky.
The BAO Schmidt telescope is equipped with
a Ford $2048 \times 2048$ Ford CCD at its main focus.
Using the 13 intermediate-band filters images of M33 obtained from the BATC
Multicolor Sky Survey, Ma et al. (2001) studied the
60 star clusters of Chandar et al. (1999a). They (Ma et al. 2001)
presented the SEDs by aperture photometry, and estimated the ages
by comparing the integrated photometric measurements with
theoretical stellar population synthesis models for these star clusters.
We can provide the accurate SEDs for these
star clusters using the multi-color photometry
of BATC.

From 35 deep $Hubble$ $Space$ $Telescope$ $(HST)$ WFPC2 fields,
Chandar, Bianchi, \& Ford (2001) again detected 102
star clustrs in M33, eighty-two of which had not previously been
detected. Using one dereddened color ($(V-I)_0$), they estimated the ages and
masses for these clusters with single stellar population models.
However, they did not give quantitative age estimates
for individual clusters due to the relatively large uncertainty
associated with age estimates from comparison of one color
with single stellar population models.

In this paper, we present the SEDs of
78 star clusters that were detected by Chandar, Bianchi, \&
Ford (2001) in M33,
and quantitatively estimate the ages for these clusters
by comparing the integrated photometric measurements with
theoretical stellar population synthesis models.

The outline of the paper is as follows.  Details of observations
and data reduction are given in section 2. In section 3, we provide
a brief description of the 
stellar population synthesis models of G. Bruzual \&
S. Charlot (1996, unpublished). The age estimates for the star clusters
are given in section 4.
The summary and discussion are presented in section 5.

\section{SAMPLE OF STAR CLASTERS, OBSERVATIONS AND DATA REDUCTION}

\subsection{Sample of Star Clusters}

The sample of star clusters in this paper is from
Chandar, Bianchi, \& Ford (2001), who used 35 deep
$Hubble$ $Space$ $Telescope$ $(HST)$ WFPC2 fields to
extend the search for star clusters in M33, and
particularly to focus on detection of older clusters.
Since these clusters cover a range of environments from
the center to the skirts, they can be used to probe the global
properties of the parent galaxy. At the same time, the accurate positions
of these star clusters
are presented in Chandar, Bianchi, \& Ford (2001).
So, we select these star clusters to be studied,
and obtain their SEDs in the 13 intermediate-band filters by aperture
photometry. The age estimates for thses star clusters are obtained
using the theoretical evolutionary
population synthesis methods. Clusters 63, 65, 66, 80, 82, 85, 102,
105, 111, 123, 134, 138, 140, 143 and 149
are not included in our sample
because of their low signal-to-noise ratio in the images
of some BATC filters. Besides, clusters 61, 70, 81, 90,
98, 104, 106, 114 and 116 are U49, M9, C20, U77,
R14, H38, H10, C38 and R12 of Christian \& Schommer (1982)
respectively,
the SEDs and ages of which were presented (Ma et al. 2002), and
are also not included in our sample. The position of cluster 85
presented by Chardar et al. (2001) may be wrong, it
should be ${\rm RA=01^h33^m14^s{\mbox{}\hspace{-0.13cm}.}28}$
decl.=30$^\circ28^{\prime}22^{\prime\prime}{\mbox{}\hspace{-0.15cm}.9}$,
and it is U137 of Christian \& Schommer (1982) (see details from
Ma et al. 2002).

Figure 1 is the image of M33 in filter BATC07 (5785{\AA}), the circles
in which indicate the positions of the sample clusters in this paper.

\begin{figure}
\figurenum{1}
\vspace{10.0cm}
\vspace{-3.5cm}
\caption{The image of M33 in filter BATC07 (5785{\AA}) and the positions
of the sample star clusters. The image size is $52'\times53'$.
The center of the image is located at
${\rm RA=01^h33^m50^s{\mbox{}\hspace{-0.13cm}.}58}$
DEC=30$^\circ39^{\prime}08^{\prime\prime}{\mbox{}\hspace{-0.15cm}.4}$
(J2000.0). North is up and east is to the left.}
\label{fig1}
\end{figure}

\subsection{Observations and Data Reduction}

The large field multi-color observations of the spiral galaxy M33 were
obtained in the BATC photometric system.
The multi-color BATC filter system, which were specifically designed to avoid
contamination from the brightest and most variable night sky emission
lines, includes 15 intermediate-band filters,
covering the total optical wavelength range from 3000 to 10000{\AA}.
The images of M33 covering
the whole optical body of M33 were accumulated in 13 intermediate band
filters with a total exposure time of about 32.75 hours from September
23, 1995 to August 28, 2000.
The dome flat-field images were taken by using a diffuse plate in
front of the correcting plate of the Schmidt telescope. For flux calibration,
the Oke-Gunn primary flux standard stars HD19445, HD84937, BD+262606
and BD+174708 were observed during photometric nights
(see details from Yan et al. 1999, Zhou et al. 2001).
{\it Column} 6 in Table 1 gives the calibration error,
in magnitude, for the standard stars in each filter. The formal
errors we obtain for these stars in the 13 BATC filters are $\la 0.02$
mag. This indicates that we can define the standard BATC system to an
accuracy of  $\la 0.02$ mag.

The data were reduced with standard procedures, including bias subtraction
and flat-fielding of the CCD images, with an automatic data reduction
software named PIPELINE I developed for the BATC multi-color sky survey
(see Ma et al. 2001, 2002 for a detail).

\subsection{Integrated Photometry}

For each star cluster, the PHOT routine in DAOPHOT (Stetson 1987, 1992)
is used to obtain magnitudes.
For avoiding contamination from nearby objects, a smaller aperture of
$6\arcsec{\mbox{}\hspace{-0.15cm}.} 8$, which corresponds
to a diameter of 4 pixels in Ford CCDs, is adopted.
Aperture corrections are computed using isolated stars.
The spectral energy distributions (SEDs) in 13 BATC filters for 78 star clusters
were obtained.  Table 2 contains the following information: {\it Column 1} is cluster
number which is taken from Chandar, Bianchi, \& Ford (2001).
{\it Column} 2 to {\it Column} 14 show the magnitudes of
different bands. Second line of each star cluster is
the uncertainties of magnitude of corresponding band.
The uncertainties for each filter are given by DAOPHOT.

\subsection{Comparison with Previous Photometry}

Using the Landolt standards, Zhou et al. (2001) presented the relationships
between the BATC intermediate-band system and $UBVRI$ broadband system
by the catalogs of Landolt (1983, 1992) and Galad\'\i-Enr\'\i quez et al. (2000).
We show the coefficients of one relationship
in equation (1).
\beq
m_V=m_{07}+(0.3233\pm0.019)(m_{06}-m_{08})+0.0590\pm0.010.
\eeq
Using equation (1), we transformed the magnitudes of
78 star clusters in BATC06, BATC07 and BATC08 bands to ones in V band.
Figure 2 plots the comparison of $V$ (BATC) photometry with previously
published measurements (Chandar, Bianchi, \& Ford 2001).
Table 3 shows this comparison. The mean $V$ magnitude difference
(this paper's values minus the values of Chandar et al. 2001) is $<\Delta V>
=0.036\pm0.042$. The uncertainties in $V$ (BATC) have been added linearly,
i.e. $\sigma_B=\sigma_{07}+0.3233(\sigma_{06}+\sigma_{08})$, to reflect the
error in the three filter measurements.
From Figure 2 and Table 3, it can be seen that
there is good agreement in the photometric measurements
between Chandar, Bianchi, \& Ford (2001) and this paper except for
clusters 115 and 127.

\begin{figure}
\figurenum{2}
\centerline{\psfig{file=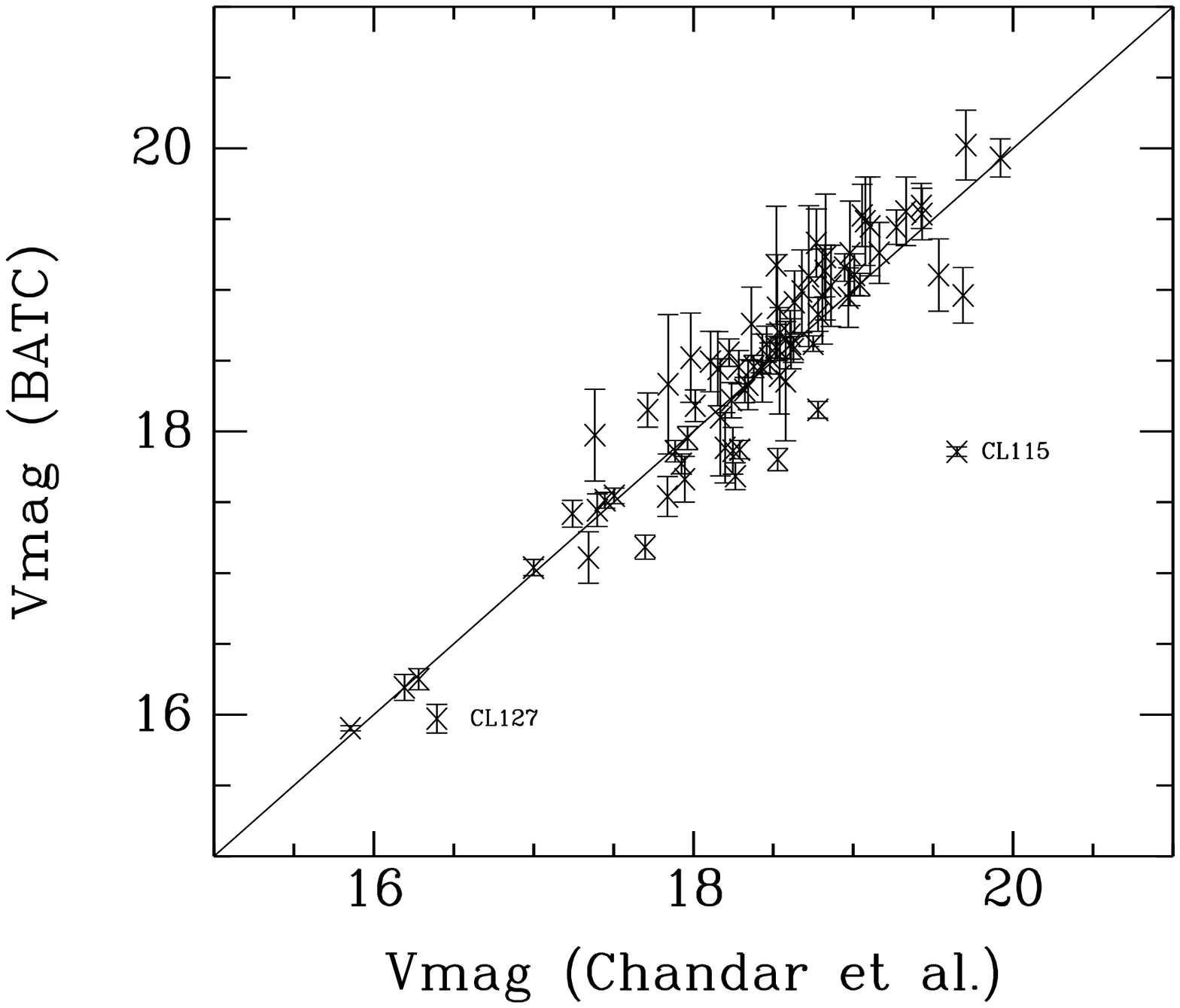,width=22.0cm,angle=-90}}
\vspace{-1.0cm}
\caption{Comparison of Cluster Photometry with Previous Measurements
($HST$)}
\label{fig1}
\end{figure}

\section{DATABASES OF SIMPLE STELLAR POPULATIONS}

Tinsley (1972) and Searle et al. (1973) did the pioneering work in
evolutionary population synthesis. This method has become
a standard technique to study the stellar populations of galaxies.
This is a result of the improvement
in the theory of the chemical evolution of galaxies, star formation,
stellar evolution and atmospheres, and of the development of synthesis
algorithms and the availability of various evolutionary synthesis models.
A comprehensive compilation of such models was presented by Leitherer et
al. (1996) and Kennicutt (1998). More widely used models are from
the Padova and Geneva group (e.g. \cite{Schaerer97}; \cite{Schaerer98};
\cite{Bressan96}; \cite{Chiosi98}), GISSEL96 (\cite{Charlot91};
\cite{Bruzual93}; G. Bruzual \& S. Charlot 1996, unpublished),
PEGASE (\cite{Fioc97}) and STARBURST99 (\cite{Leitherer99}).

A simple stellar population (SSP) is defined as a single generation
of coeval stars with fixed parameters such as metallicity, initial
mass function, etc. (\cite{Buzzoni97}).
SSPs are the basic building blocks of synthetic spectra
of galaxies that can be used to infer the formation and subsequent
evolution of the parent galaxies (\cite{Jab96}).
They are modeled by a collection of stellar evolutionary tracks with
different masses and initial chemical compositions, supplemented
with a library of stellar spectra for stars at different evolutionary
stages in evolution synthesis models. In this paper, we use  the SSPs
of Galaxy Isochrone Synthesis Spectra Evolution Library
(hereafter GSSP; G. Bruzual \& S. Charlot 1996, unpublished) to estimate the ages of
the sample clusters, since they are simple and reasonably well understood.
 
\subsection{Spectral Energy Distribution of GSSPs}

Charlot \& Bruzual (1991) developed a model of
stellar population synthesis. In this model, the
population synthesis method can be used to determine
the distribution of stars in the
theoretical color-magnitude diagram (CMD) for
any stellar system. Bruzual \& Charlot (1993)
presented ``isochrone synthesis'' as a natural and
reliable approach to model the evolution of stellar
populations in star clusters and galaxies. With
this isochrone synthesis algorithm, Bruzual \& Charlot (1993)
computed the spectral energy distributions of stellar
populations with solar metallicity.
G. Bruzual \& S. Charlot (1996, unpublished) improved the Bruzual \& Charlot
(1993) evolutionary population synthesis models. The updated version
provides the evolution of the spectrophotometric properties for a wide
range of stellar metallicity, which are $Z=0.0004, 0.004, 0.008,
0.02, 0.05,$ and $0.1$ (see Ma et al. 2001, 2002 for a detail).

\subsection{Integrated Colors of GSSPs}

Kong et al. (2000) have obtained the age, metallicity, and interstellar-medium
reddening distribution for M81.
They found the best match between the intrinsic colors
and the predictions of GSSP for each cell of M81.
To estimate the ages for the sample clusters in this paper,
we follow the method of Kong et al. (2000).
As we know, the observational
data are integrated luminosity. So, we need to convolve
the SED of GSSP with BATC filter profiles to obtain the optical
and near-infrared integrated luminosity for comparisons (Kong et al. 2000).
The integrated luminosity
$L_{\lambda_i}(t,Z)$ of the $i$th BATC filter can be calculated with
\beq 
L_{\lambda_i}(t,Z) =\frac{\int
F_{\lambda}(t,Z)\varphi_i(\lambda)d\lambda} {\int
\varphi_i(\lambda)d\lambda},
\eeq
where $F_{\lambda}(t,Z)$ is the spectral energy distribution of
the GSSP of metallicity $Z$ at age $t$, $\varphi_i(\lambda)$ is the
response functions of the $i$th filter of the BATC filter system
($i=3, 4, \cdot\cdot\cdot, 15$),
respectively.
For avoiding to use the parameters that are
denpantant on the distance.
We calculate the integrated colors of a GSSP relative to
the BATC filter BATC08 ($\lambda=6075${\AA}):
\beq 
\label{color}
C_{\lambda_i}(t,Z)={L_{\lambda_i}(t,Z)}/{L_{6075}(t,Z)}.  
\eeq
As a result, we obtained the intermediate-band colors
of a GSSP for 6 metallicities from Z=0.0004 to Z=0.1 using
equations (2) and (3).

\section{AGE ESTIMATES}

In order to obtain intrinsic colors of 78 clusters and hence
accurate ages,
the photometric measurements must be dereddened.
As Chandar, Bianchi, \& Ford (2001) did, we adopted
$E(B-V)=0.10$. Besides,
we adopted the extinction curve presented by Zombeck (1990).
An extinction correction $A_{\lambda}=R_{\lambda}E(B-V)$ was
applied, here $R_{\lambda}$ is obtained by interpolating
using the data of Zombeck (1990).

Since we model the stellar populations of the star clusters
by SSPs, the intrinsic colors for
each star cluster are determined by two parameters: age, and metallicity.
We will determine the ages and best-fitted models
of metallicity for these
star clusters simultaneously by a least square method.
The age and best-fitted model of metallicity
are found by minimizing the difference between the intrinsic and
integrated colors of GSSP:

\beq 
R^2(n,t,Z)=\sum_{i=3}^{15}[C_{\lambda_i}^{\rm
intr}(n)-C_{\lambda_i}^ {\rm ssp}(t, Z)]^2, 
\eeq
where $C_{\lambda_i}^{\rm ssp}(t, Z)$ represents the integrated color in
the $i$th filter of a SSP at age $t$ in the model of metallicity $Z$,
and $C_{\lambda_i}^{\rm intr}(n)$ is the intrinsic
integrated color for nth star cluster.
Using the stellar evolutionary models (Bertelli et al. 1994)
and published line indices of 22 M33 older clusters,
Chandar, Bianchi, \&
Ford (1999b) narrowed the range of cluster metallicities (Z)
to be from $\sim 0.0002$ to 0.03. So,
we only select the models of three metallicities, 0.0004, 0.004 and 0.02 of
GSSP.

Figure 3 shows the map of the best fit of the integrated color
of a SSP with the intrinsic integrated color for 78 star clusters,
and Table 4 presents the best-fitted models of metallicities and ages for
these star clusters.
In Figure 3, the thick line represents the integrated
color of a SSP of GSSP, and filled circle represents the intrinsic
integrated color of a star cluster.
From this figure, we see that clusters 83, 88 and 148
have strong emission lines. In the process of fitting, we did
not use the strong emission lines.

\begin{figure}
\figurenum{3}
\centerline{\psfig{file=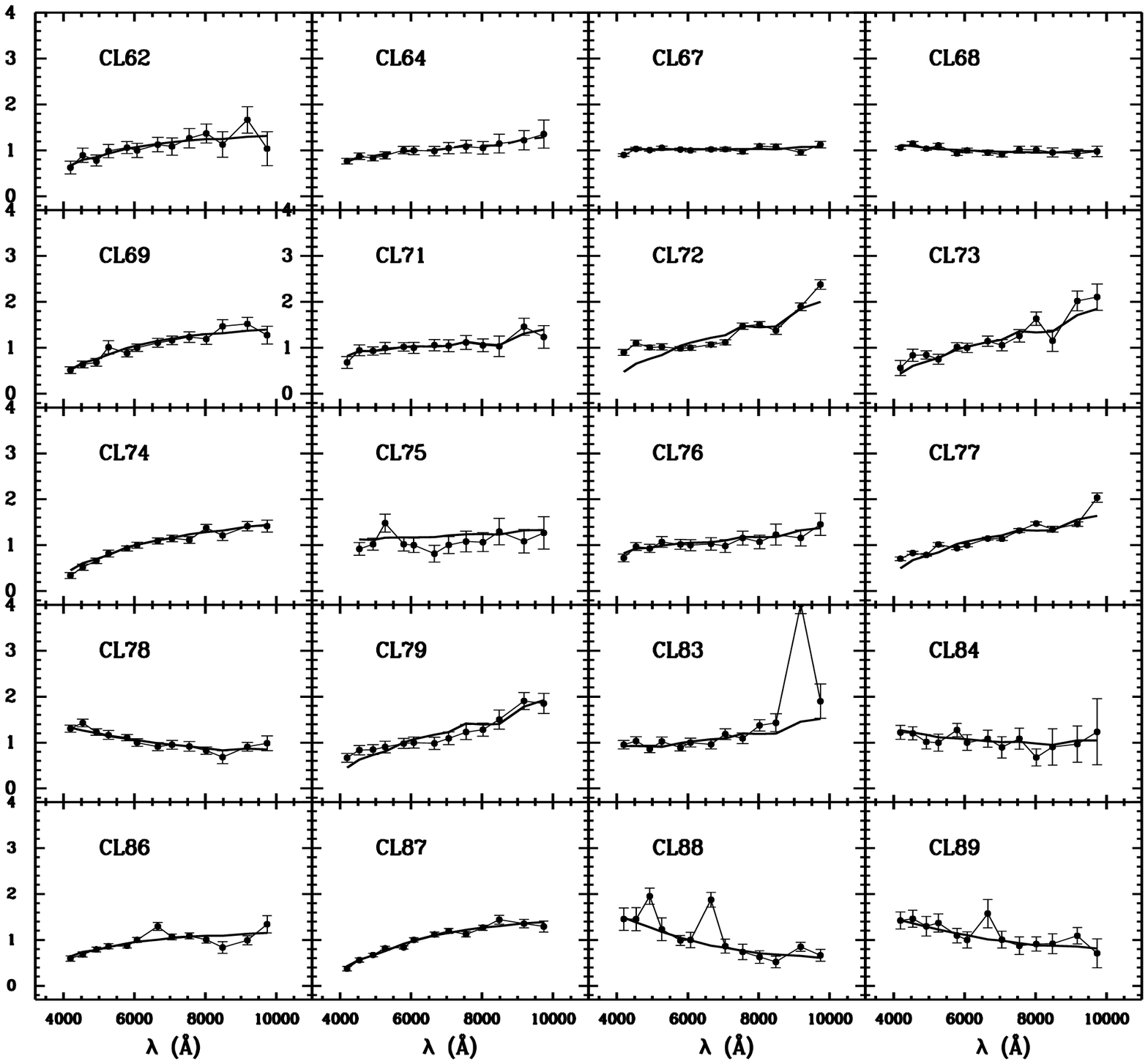,width=22.0cm,angle=270}}
\caption{Map of the best fit of the integrated color
of a SSP with intrinsic integrated color for 78 star clusters.
Thick line represents the integrated color of a SSP, and
filled circle represents the intrinsic integrated color of a star cluster.}
\end{figure}

\begin{figure}
\figurenum{3}
\centerline{\psfig{file=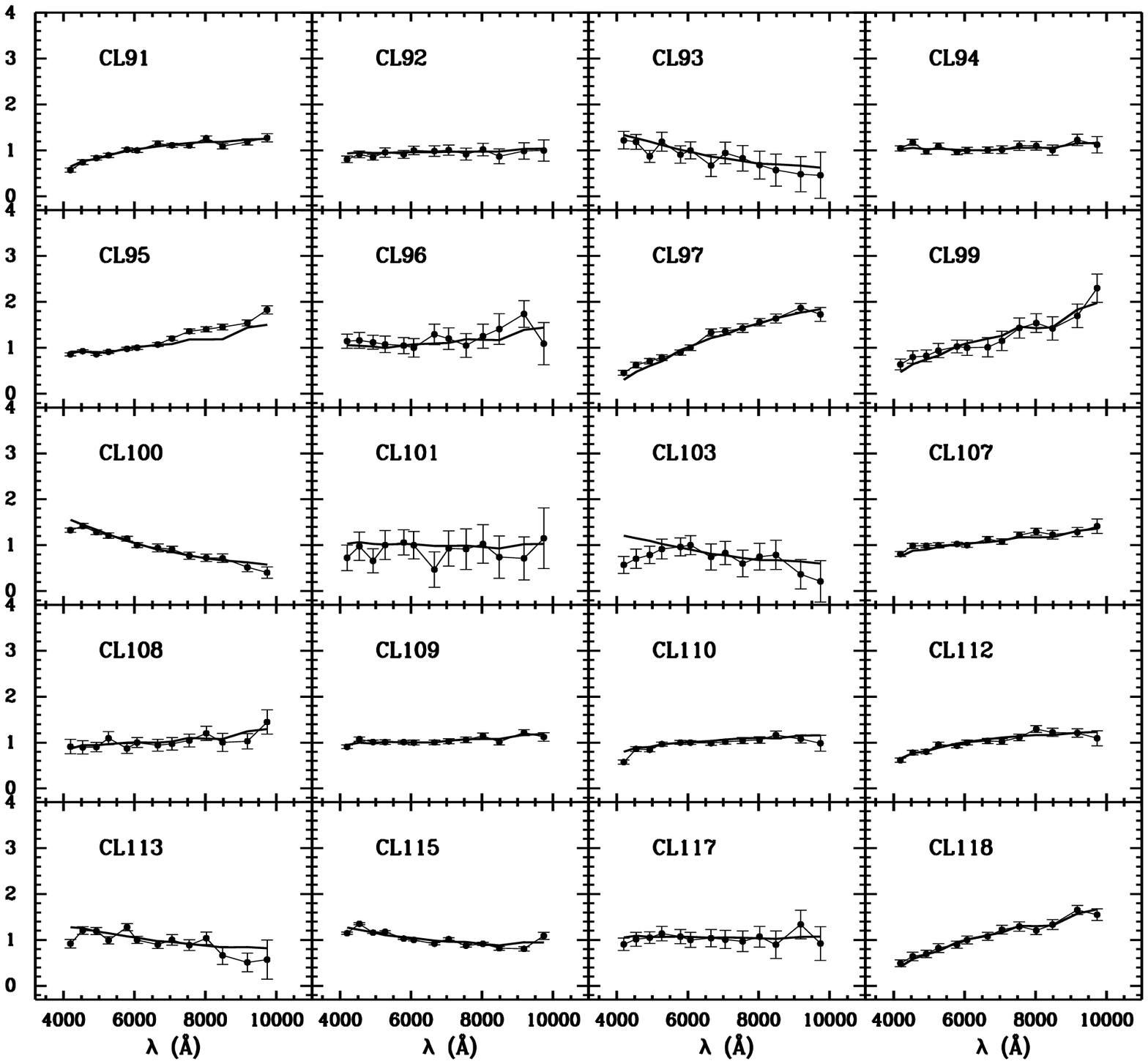,width=22.0cm,angle=270}}
\caption{Continued}
\end{figure}

\begin{figure}
\figurenum{3}
\centerline{\psfig{file=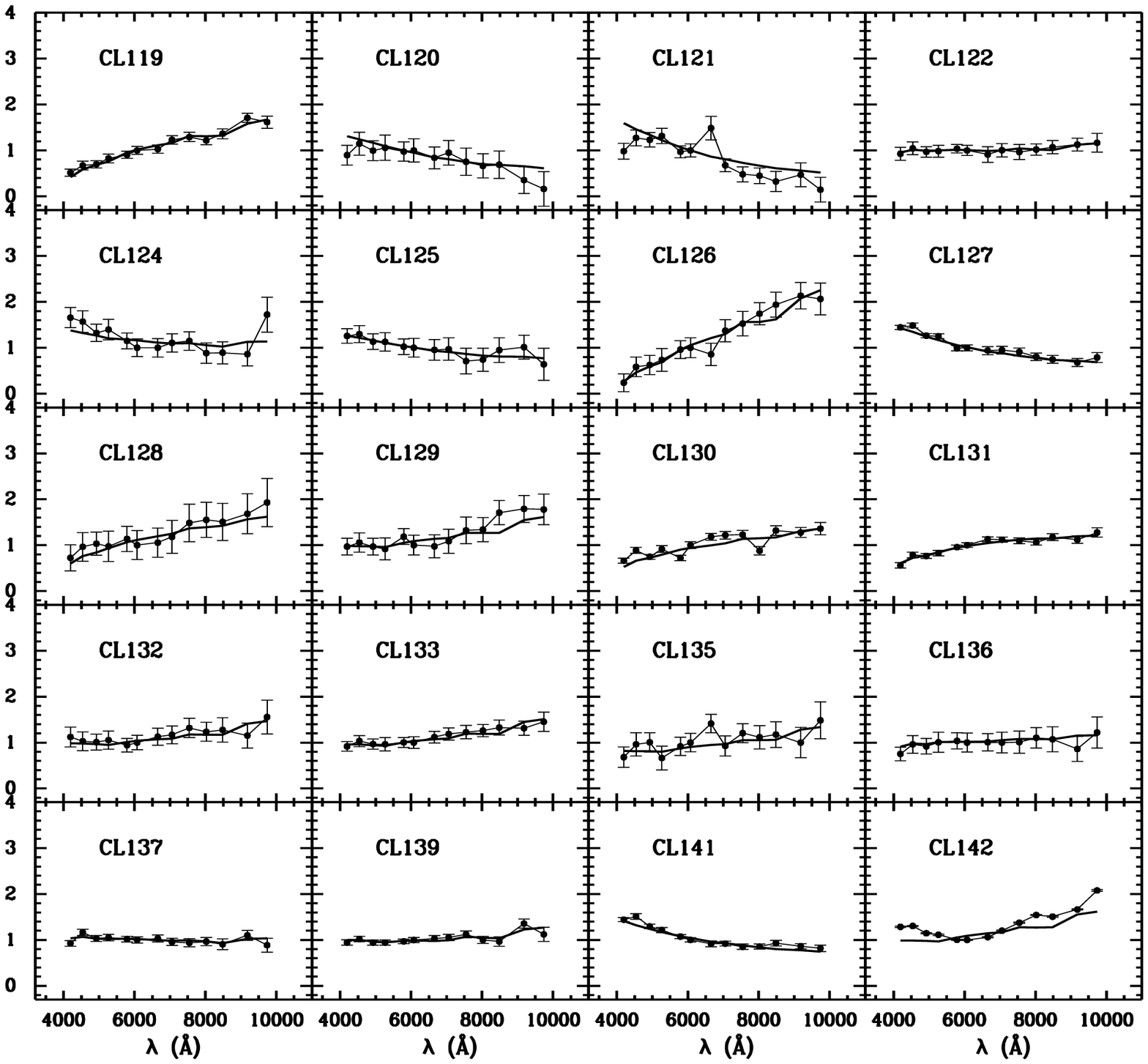,width=22.0cm,angle=270}}
\caption{Continued}
\end{figure}

\begin{figure}
\figurenum{3}
\centerline{\psfig{file=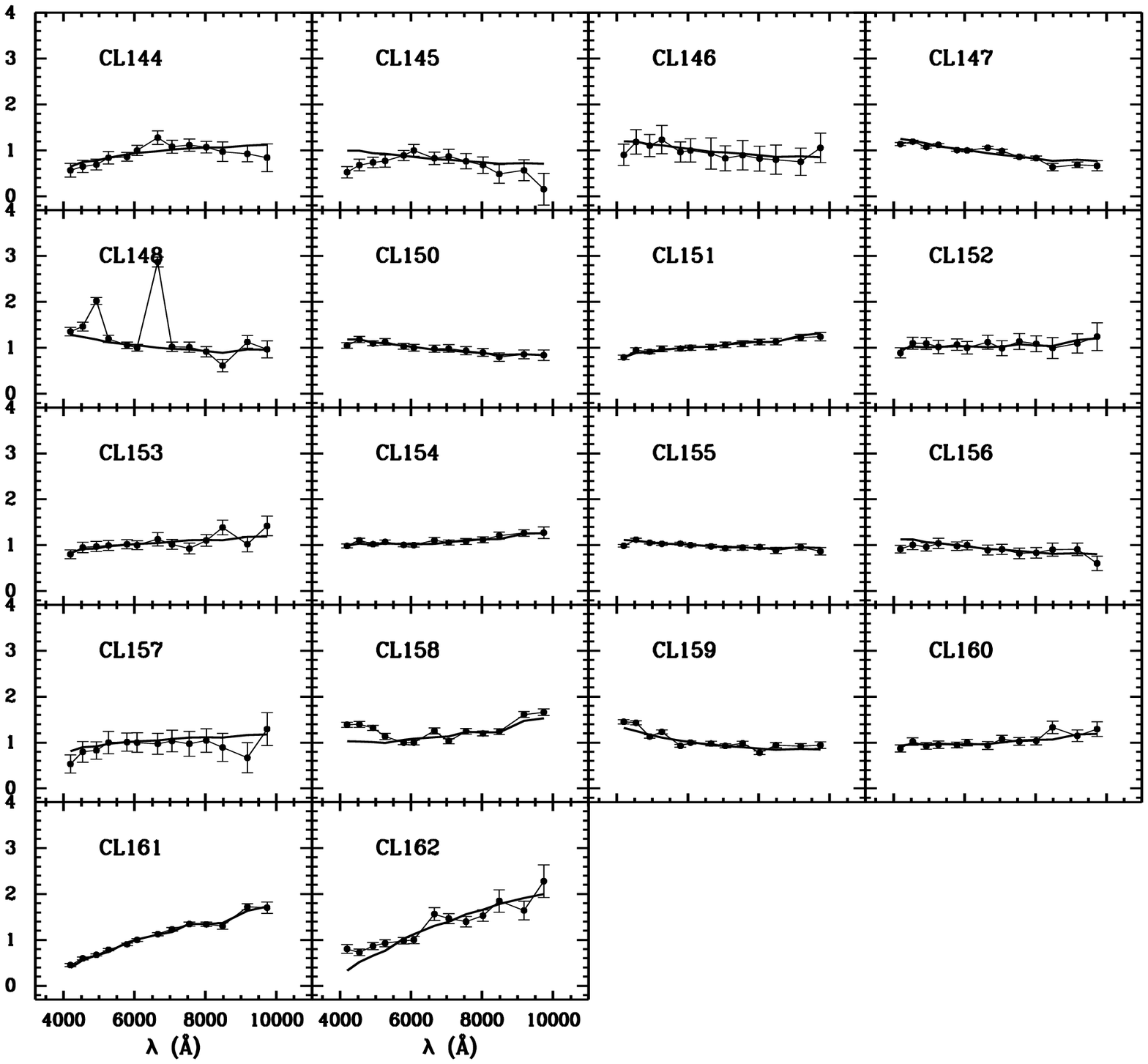,width=22.0cm,angle=270}}
\caption{Continued}
\end{figure}

Figure 4 presents a histogram of cluster ages. The results show that,
in general, M33 clusters have been forming continuously, with
ages of $\sim 3 \times 10^{6}$ -- $10^{10}$ years.
This conclusion confirms the results of Chandar, Bianchi, \& Fort (2001).
There exist three groups of clusters that formed with three models of
metalicities, $Z=0.02, 0.004$, and $0.0004$. In different models of metallicities,
the distribution of cluster ages is a little different, too.
In the model of $Z=0.02$, the ages of most clusters
are younger than $\sim 10^{9}$ years, and there are two peaks
at $\sim 10^{7}$ and $\sim 10^{9}$ years.
In the model of $Z=0.004$,
the clusters formed from $\sim 3 \times 10^{6}$ -- $10^{10}$ years,
and the distribution of ages is more homogeneous than in the other
two models.
In the model of $Z=0.0004$, the most clusters formed from
$\sim 10^{8}$ -- $10^{10}$ years.
Clusters 97, 106 and 162 have derived ages consistent with that of the globular
clusters of the Milky Way, $\sim 1.5\times 10^{10}$ years. This result is
also consistent with
that found by Chandar, Bianchi, \& Fort (1999b) and Ma et al. (2001),
who presented clusters 11, 28, 29 and 57 to be as old as
$\sim 1.5\times 10^{10}$ years.

In this section, we estimate
the ages of our sample clusters by comparing the photometry
of each object with the theoretical stellar population
synthesis models for different values of metallicity.
However, we want to emphasize that, for clusters older than several
$10^8$ years, the age/metallicity degeneracy becomes
pronounced. In this case, we only mean that in some model of
metallicity, the intrinsic integrated color of a cluster can do
the best fit with the integrated color of a SSP at some age.
Besides, the uncertainties in the age estimates arising from
photometric uncertainties are 0.2 or so, i.e,
$\rm{age}\pm 0.2\times\rm{age}~[\log \rm{yr}]$.

\begin{figure}
\figurenum{4}
\centerline{\psfig{file=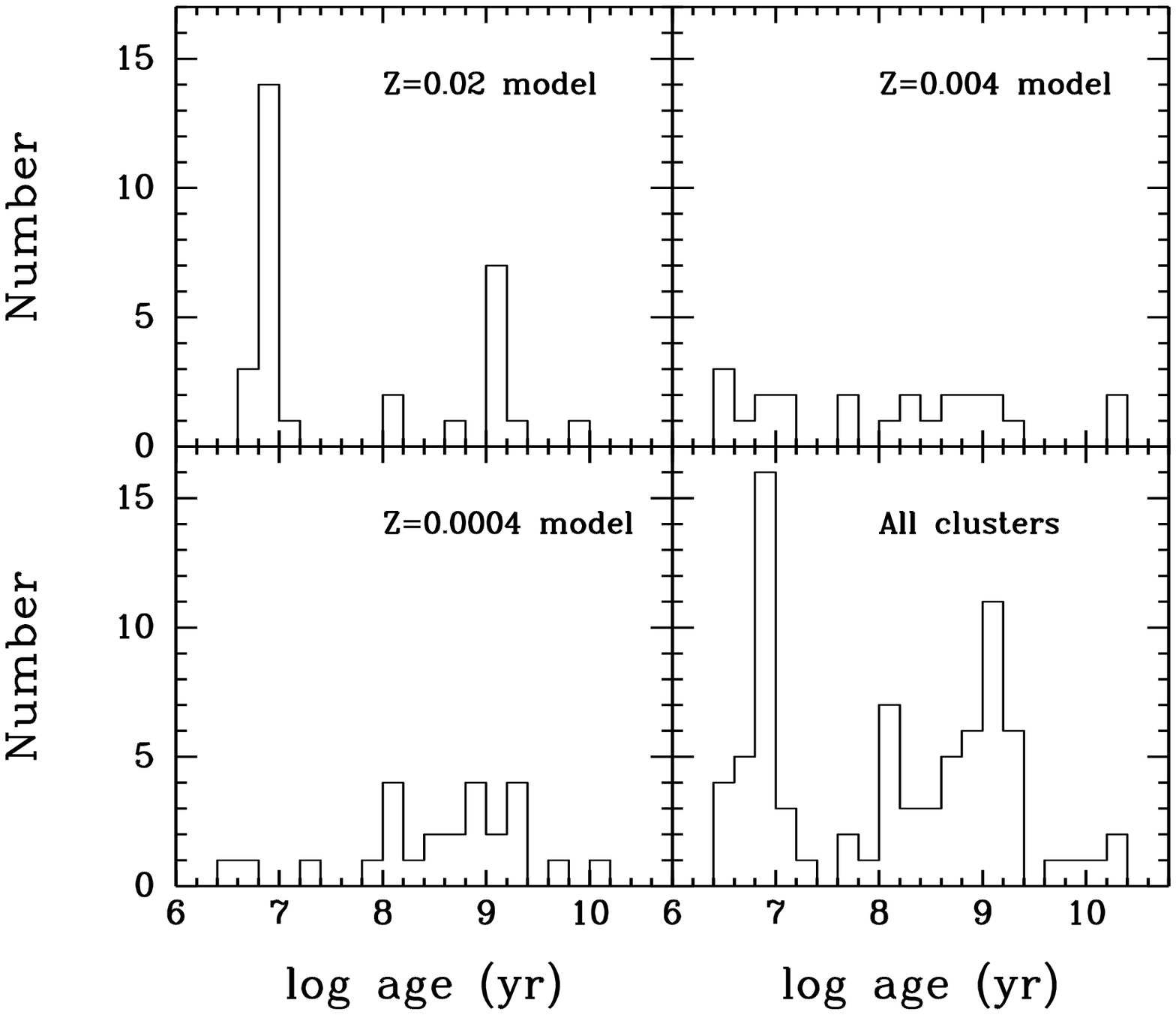,width=24.0cm,angle=-90}}
\vspace{-2.0cm}
\caption{Histogram of M33 cluster ages}
\label{fig4}
\end{figure}

\section{SUMMARY AND DISCUSSION}

In this paper, we have, for the first time, obtained the SEDs
of 78 star clusters of M33 in 13 intermediate colors with
the BAO 60/90 cm Schmidt telescope.
Below, we summarize our main conclusions.

1. Using the images obtained with the Beijing Astronomical
Observatory $60/90$ cm Schmidt Telescope in 13 intermediate-band filters
from 3800 to 10000{\AA}, we
obtained the spectral energy
distributions (SEDs) of 78 star clusters
that were detected by Chandar, Bianchi, \&
Ford (2001).

2. By comparing the integrated photometric measurements with
theoretical stellar population synthesis models,
we find that clusters formed continuously in
M33 from $\sim 3\times10^6$ -- $10^{10}$ years.
The results also show that, there are two peaks at
$\sim 8\times10^6$ and $\sim 10^9$ years.

Chandar et al. (1999a, 1999b) estimated ages for 60 star clusters
in M33 by comparing the photometric measurements to
integrated color from theoretical models by Bertelli et al. (1994).
Their results showed that, the integrated colors of star clusters
depend mostly on age, with a secondary dependence on chemical composition.
So, we can estimate ages of clusters, but cannot determine
metallicities of clusters exactly. As Chandar, Bianchi, \& Ford (1999b, 1999c,
2001) did, we also estimated the ages of our sample clusters by comparing
the photometry of each object with models for different values of metallicity.
Although we presented the metallity of each cluster in Table 4,
we only mean that, in this model of metallicity, the intrinsic integrated color
of each cluster can do the best fit with the integrated color of a SSP.

With spectrophotometry,
Christian \& Schommer (1983) obtained the ages of the star clusters
in M33 to be $\sim 10^7$ -- $10^{10}$ years.
Using the integrated
$UBV$ photometry and IUE $\lambda\lambda1200-3000$ \AA~ spectra, Ciani,
D'Odorico, \& Benvenuti (1984) studied the minuscule ``bulge''
population of M33 and found that, a multigeneration model,
where a young component (age $\sim 10^7$ years) and an old,
metal-poor one (age $\sim 5\times10^9$ years) are superposed,
gives the best fit to the observed data. Schmidt, Bica, \&
Alloin (1990) applied a population synthesis method which uses a
star cluster spectral library and a grid of the star
cluster spectral properties as a function of age and metallicity
(Bica \& Alloin 1986a, b; 1987),
to the blueish nucleus of M33, and gave an age of less than
$5\times10^8$ years for the dominant blue bulge population.
From the histogram of ages in this paper, we can see that
some old clusters in our sample appear to be coeval with the
old population of the bulge.

\acknowledgments
We would like to thank the anonymous referee for his/her
insightful comments and suggestions that improved this paper.
We are grateful to the Padova group for providing us with a set of
theoretical isochrones and SSPs. We also thank G. Bruzual and
S. Charlot for sending us their latest calculations of SSPs and
for explanations of their code.
The work is supported partly by the National Sciences
Foundation under  the contract  No.19833020 and No.19503003.
The BATC Survey is supported by the
Chinese Academy of Sciences, the Chinese National Natural Science
Foundation and the Chinese State Committee of Sciences and
Technology.
The project is also supported in part
by the National Science Foundation (grant INT 93-01805) and
by Arizona State University, the University of Arizona and Western
Connecticut State University.

\clearpage
\setcounter{table}{0}
\begin{table}[htb]
\caption[]{Parameters of the BATC Filters and Statistics of Observations}
\vspace {0.5cm}
\begin{tabular}{cccccc}
\hline
\hline
 No. & Name& cw\tablenotemark{a}~~(\AA)& Exp. (hr)&  N.img\tablenotemark{b}
 & rms\tablenotemark{c} \\
\hline
1  & BATC03& 4210   & 00:55& 04 &0.024\\
2  & BATC04& 4546   & 01:05& 04 &0.023\\
3  & BATC05& 4872   & 03:55& 19 &0.017\\
4  & BATC06& 5250   & 03:19& 15 &0.006\\
5  & BATC07& 5785   & 04:38& 17 &0.011\\
6  & BATC08& 6075   & 01:26& 08 &0.016\\
7  & BATC09& 6710   & 01:09& 08 &0.006\\
8  & BATC10& 7010   & 01:41& 08 &0.005\\
9  & BATC11& 7530   & 02:07& 10 &0.017\\
10 & BATC12& 8000   & 03:00& 11 &0.003\\
11 & BATC13& 8510   & 03:15& 11 &0.005\\
12 & BATC14& 9170   & 01:15& 05 &0.011\\
13 & BATC15& 9720   & 05:00& 26 &0.009\\
\hline
\end{tabular}\\
\tablenotetext{\rm a}{Central wavelength for each BATC filter}
\tablenotetext{\rm b}{Image numbers for each BATC filter}
\tablenotetext{\rm c}{Calibration error, in magnitude, for each filter
as obtained from the standard stars}
\end{table}

\clearpage
{\small
\setcounter{table}{1}
\begin{table}[ht]
\caption{SEDs of 78 Star Clusters}
\vspace {0.3cm}
\begin{tabular}{cccccccccccccc}
\hline
\hline
 No. & 03  &  04 &  05 &  06 &  07 &  08 &  09 &  10 &  11 &  12 &  13 &  14 &  15\\
(1)    & (2) & (3) & (4) & (5) & (6) & (7) & (8) & (9) & (10) & (11) & (12) & (13) & (14)\\
\hline
      62 & 19.970 & 19.551 & 19.672 & 19.377 & 19.248 & 19.301 & 19.142 & 19.172 & 18.984 & 18.881 & 19.070 & 18.627 & 19.131\\
   & 0.238 &  0.186 &  0.168 &  0.162 &  0.133 &  0.168 &  0.151 &  0.188 &  0.180 &  0.166 &  0.268 &  0.187 &  0.389 \\
      64 & 19.433 & 19.262 & 19.282 & 19.171 & 18.988 & 18.984 & 18.969 & 18.891 & 18.837 & 18.848 & 18.735 & 18.646 & 18.524\\
   & 0.089 &  0.087 &  0.082 &  0.094 &  0.084 &  0.100 &  0.111 &  0.127 &  0.137 &  0.140 &  0.192 &  0.187 &  0.244 \\
      67 & 17.742 & 17.563 & 17.559 & 17.470 & 17.456 & 17.469 & 17.414 & 17.402 & 17.435 & 17.293 & 17.288 & 17.396 & 17.208\\
   & 0.034 &  0.033 &  0.032 &  0.039 &  0.032 &  0.034 &  0.037 &  0.044 &  0.052 &  0.049 &  0.060 &  0.068 &  0.068 \\
      68 & 17.925 & 17.801 & 17.883 & 17.773 & 17.910 & 17.824 & 17.849 & 17.879 & 17.747 & 17.729 & 17.774 & 17.784 & 17.718\\
   & 0.032 &  0.036 &  0.035 &  0.043 &  0.051 &  0.052 &  0.066 &  0.071 &  0.083 &  0.084 &  0.113 &  0.115 &  0.125 \\
      69 & 19.363 & 19.100 & 18.992 & 18.525 & 18.623 & 18.478 & 18.336 & 18.276 & 18.191 & 18.208 & 17.961 & 17.903 & 18.083\\
   & 0.154 &  0.125 &  0.135 &  0.155 &  0.100 &  0.088 &  0.093 &  0.087 &  0.099 &  0.103 &  0.106 &  0.102 &  0.163 \\
      71 & 19.640 & 19.258 & 19.241 & 19.131 & 19.054 & 19.065 & 18.974 & 18.977 & 18.887 & 18.928 & 18.931 & 18.535 & 18.705\\
   & 0.205 &  0.135 &  0.105 &  0.134 &  0.097 &  0.131 &  0.125 &  0.134 &  0.146 &  0.145 &  0.233 &  0.138 &  0.220 \\
      72 & 18.468 & 18.216 & 18.284 & 18.230 & 18.222 & 18.193 & 18.091 & 18.031 & 17.715 & 17.667 & 17.745 & 17.380 & 17.122\\
   & 0.080 &  0.062 &  0.055 &  0.070 &  0.048 &  0.058 &  0.052 &  0.056 &  0.051 &  0.046 &  0.067 &  0.047 &  0.048 \\
      73 & 20.156 & 19.692 & 19.652 & 19.739 & 19.357 & 19.368 & 19.191 & 19.269 & 19.056 & 18.755 & 19.113 & 18.484 & 18.430\\
   & 0.319 &  0.171 &  0.112 &  0.155 &  0.097 &  0.110 &  0.104 &  0.129 &  0.118 &  0.097 &  0.220 &  0.117 &  0.149 \\
      74 & 19.926 & 19.425 & 19.138 & 18.868 & 18.678 & 18.589 & 18.465 & 18.399 & 18.410 & 18.164 & 18.281 & 18.094 & 18.081\\
   & 0.201 &  0.142 &  0.095 &  0.100 &  0.064 &  0.069 &  0.064 &  0.068 &  0.072 &  0.061 &  0.098 &  0.076 &  0.101 \\
      75 & 19.782 & 19.400 & 19.256 & 18.811 & 19.164 & 19.177 & 19.370 & 19.133 & 19.033 & 19.028 & 18.796 & 18.969 & 18.790\\
   & 0.190 &  0.164 &  0.139 &  0.143 &  0.155 &  0.169 &  0.246 &  0.199 &  0.228 &  0.211 &  0.240 &  0.256 &  0.303 \\
      76 & 19.427 & 19.077 & 19.093 & 18.900 & 18.906 & 18.913 & 18.852 & 18.893 & 18.697 & 18.758 & 18.588 & 18.634 & 18.377\\
   & 0.127 &  0.098 &  0.105 &  0.122 &  0.115 &  0.129 &  0.145 &  0.154 &  0.144 &  0.151 &  0.202 &  0.161 &  0.180 \\
      77 & 18.534 & 18.327 & 18.353 & 18.037 & 18.080 & 17.995 & 17.819 & 17.816 & 17.635 & 17.497 & 17.572 & 17.464 & 17.093\\
   & 0.061 &  0.045 &  0.039 &  0.046 &  0.032 &  0.036 &  0.030 &  0.036 &  0.035 &  0.028 &  0.049 &  0.037 &  0.054 \\
      78 & 18.169 & 18.041 & 18.174 & 18.191 & 18.196 & 18.297 & 18.368 & 18.312 & 18.333 & 18.426 & 18.613 & 18.282 & 18.184\\
   & 0.065 &  0.064 &  0.064 &  0.090 &  0.066 &  0.092 &  0.100 &  0.110 &  0.124 &  0.104 &  0.225 &  0.114 &  0.175 \\
      79 & 19.398 & 19.123 & 19.082 & 18.970 & 18.831 & 18.799 & 18.789 & 18.665 & 18.516 & 18.452 & 18.258 & 17.980 & 18.000\\
   & 0.153 &  0.132 &  0.139 &  0.152 &  0.118 &  0.133 &  0.148 &  0.138 &  0.143 &  0.120 &  0.149 &  0.105 &  0.130 \\
      83 & 19.602 & 19.485 & 19.652 & 19.422 & 19.523 & 19.388 & 19.405 & 19.166 & 19.233 & 18.962 & 18.900 & 17.760 & 18.561\\
   & 0.106 &  0.099 &  0.100 &  0.104 &  0.092 &  0.103 &  0.109 &  0.108 &  0.114 &  0.097 &  0.148 &  0.055 &  0.214 \\
      84 & 20.139 & 20.126 & 20.281 & 20.259 & 19.944 & 20.196 & 20.079 & 20.275 & 20.047 & 20.539 & 20.205 & 20.112 & 19.839\\
   & 0.134 &  0.129 &  0.164 &  0.202 &  0.123 &  0.183 &  0.185 &  0.279 &  0.228 &  0.296 &  0.478 &  0.446 &  0.634 \\
      86 & 19.590 & 19.409 & 19.220 & 19.089 & 19.027 & 18.867 & 18.556 & 18.758 & 18.716 & 18.780 & 18.963 & 18.753 & 18.418\\
   & 0.107 &  0.087 &  0.073 &  0.070 &  0.056 &  0.056 &  0.072 &  0.065 &  0.067 &  0.083 &  0.163 &  0.108 &  0.153 \\
      87 & 19.913 & 19.440 & 19.219 & 18.963 & 18.888 & 18.684 & 18.533 & 18.450 & 18.497 & 18.348 & 18.188 & 18.234 & 18.276\\
   & 0.143 &  0.090 &  0.060 &  0.057 &  0.045 &  0.051 &  0.051 &  0.052 &  0.055 &  0.047 &  0.072 &  0.075 &  0.102 \\ 
\end{tabular}
\end{table}
}

\clearpage
{\small
\setcounter{table}{1}
\begin{table}[htb]
\caption{Continued}
\vspace {0.3cm}
\begin{tabular}{cccccccccccccc}
\hline
\hline
 No. & 03  &  04 &  05 &  06 &  07 &  08 &  09 &  10 &  11 &  12 &  13 &  14 &  15\\
(1)    & (2) & (3) & (4) & (5) & (6) & (7) & (8) & (9) & (10) & (11) & (12) & (13) & (14)\\
\hline
      88 & 17.608 & 17.579 & 17.228 & 17.690 & 17.878 & 17.855 & 17.143 & 17.973 & 18.123 & 18.281 & 18.468 & 17.912 & 18.170\\
   & 0.182 &  0.185 &  0.098 &  0.218 &  0.119 &  0.181 &  0.094 &  0.190 &  0.244 &  0.232 &  0.266 &  0.128 &  0.209 \\
      89 & 18.292 & 18.232 & 18.331 & 18.232 & 18.425 & 18.514 & 17.991 & 18.468 & 18.601 & 18.534 & 18.508 & 18.300 & 18.758\\
   & 0.142 &  0.136 &  0.177 &  0.158 &  0.156 &  0.190 &  0.213 &  0.195 &  0.238 &  0.186 &  0.256 &  0.179 &  0.484 \\
      91 & 18.517 & 18.203 & 18.047 & 17.933 & 17.743 & 17.750 & 17.569 & 17.596 & 17.579 & 17.416 & 17.556 & 17.447 & 17.358\\
   & 0.092 &  0.074 &  0.060 &  0.056 &  0.042 &  0.045 &  0.051 &  0.042 &  0.046 &  0.043 &  0.059 &  0.062 &  0.076 \\
      92 & 18.876 & 18.707 & 18.755 & 18.590 & 18.589 & 18.481 & 18.469 & 18.433 & 18.515 & 18.378 & 18.535 & 18.376 & 18.353\\
   & 0.095 &  0.089 &  0.079 &  0.116 &  0.102 &  0.098 &  0.125 &  0.121 &  0.154 &  0.145 &  0.200 &  0.197 &  0.250 \\
      93 & 19.262 & 19.262 & 19.566 & 19.190 & 19.431 & 19.318 & 19.722 & 19.339 & 19.461 & 19.659 & 19.826 & 19.992 & 20.034\\
   & 0.171 &  0.147 &  0.162 &  0.189 &  0.223 &  0.203 &  0.387 &  0.269 &  0.363 &  0.488 &  0.660 &  0.867 &  1.192 \\
      94 & 18.533 & 18.374 & 18.544 & 18.382 & 18.469 & 18.420 & 18.382 & 18.364 & 18.255 & 18.239 & 18.313 & 18.079 & 18.165\\
   & 0.059 &  0.060 &  0.064 &  0.068 &  0.062 &  0.076 &  0.074 &  0.090 &  0.099 &  0.092 &  0.119 &  0.109 &  0.174 \\
      95 & 18.055 & 17.941 & 17.991 & 17.887 & 17.762 & 17.728 & 17.622 & 17.487 & 17.335 & 17.280 & 17.223 & 17.139 & 16.944\\
   & 0.040 &  0.038 &  0.036 &  0.041 &  0.038 &  0.041 &  0.042 &  0.039 &  0.041 &  0.042 &  0.046 &  0.045 &  0.054 \\
      96 & 19.486 & 19.442 & 19.451 & 19.457 & 19.432 & 19.473 & 19.164 & 19.234 & 19.361 & 19.148 & 19.002 & 18.754 & 19.249\\
   & 0.145 &  0.163 &  0.147 &  0.181 &  0.183 &  0.219 &  0.187 &  0.212 &  0.269 &  0.228 &  0.257 &  0.181 &  0.459 \\
      97 & 19.183 & 18.805 & 18.641 & 18.486 & 18.293 & 18.168 & 17.825 & 17.793 & 17.723 & 17.607 & 17.532 & 17.369 & 17.444\\
   & 0.125 &  0.105 &  0.091 &  0.086 &  0.067 &  0.068 &  0.063 &  0.059 &  0.069 &  0.057 &  0.065 &  0.056 &  0.094 \\ 
      99 & 19.015 & 18.739 & 18.673 & 18.494 & 18.342 & 18.363 & 18.321 & 18.171 & 17.915 & 17.816 & 17.880 & 17.670 & 17.328\\
   & 0.204 &  0.185 &  0.167 &  0.179 &  0.146 &  0.181 &  0.226 &  0.199 &  0.167 &  0.145 &  0.192 &  0.164 &  0.146 \\
     100 & 17.161 & 17.057 & 17.137 & 17.160 & 17.172 & 17.307 & 17.342 & 17.371 & 17.529 & 17.575 & 17.569 & 17.901 & 18.167\\
   & 0.036 &  0.039 &  0.043 &  0.052 &  0.049 &  0.059 &  0.099 &  0.085 &  0.113 &  0.120 &  0.141 &  0.207 &  0.339 \\
     101 & 19.580 & 19.227 & 19.624 & 19.130 & 19.019 & 19.071 & 19.863 & 19.110 & 19.108 & 18.962 & 19.296 & 19.319 & 18.789\\
   & 0.414 &  0.344 &  0.431 &  0.343 &  0.282 &  0.326 &  0.899 &  0.448 &  0.524 &  0.442 &  0.677 &  0.715 &  0.623 \\
     103 & 19.482 & 19.223 & 19.071 & 18.870 & 18.764 & 18.713 & 19.007 & 18.876 & 19.205 & 18.948 & 18.872 & 19.680 & 20.271\\
   & 0.354 &  0.321 &  0.257 &  0.259 &  0.218 &  0.225 &  0.418 &  0.333 &  0.528 &  0.422 &  0.438 &  0.945 &  2.321 \\
     107 & 18.788 & 18.541 & 18.507 & 18.462 & 18.378 & 18.395 & 18.229 & 18.273 & 18.119 & 18.037 & 18.070 & 18.003 & 17.890\\
   & 0.070 &  0.059 &  0.049 &  0.060 &  0.044 &  0.049 &  0.053 &  0.058 &  0.059 &  0.061 &  0.084 &  0.084 &  0.119 \\
     108 & 19.316 & 19.304 & 19.268 & 19.017 & 19.215 & 19.055 & 19.091 & 19.046 & 18.948 & 18.775 & 18.952 & 18.904 & 18.524\\
   & 0.188 &  0.178 &  0.124 &  0.140 &  0.133 &  0.118 &  0.151 &  0.152 &  0.146 &  0.140 &  0.212 &  0.174 &  0.200 \\
     109 & 17.991 & 17.778 & 17.812 & 17.777 & 17.727 & 17.726 & 17.692 & 17.658 & 17.605 & 17.498 & 17.608 & 17.392 & 17.472\\
   & 0.049 &  0.044 &  0.042 &  0.054 &  0.044 &  0.054 &  0.053 &  0.058 &  0.063 &  0.056 &  0.074 &  0.059 &  0.090 \\
     110 & 19.262 & 18.791 & 18.784 & 18.594 & 18.506 & 18.497 & 18.480 & 18.440 & 18.391 & 18.358 & 18.232 & 18.295 & 18.385\\
   & 0.087 &  0.067 &  0.057 &  0.048 &  0.048 &  0.046 &  0.052 &  0.055 &  0.066 &  0.073 &  0.083 &  0.086 &  0.190 \\
     112 & 19.082 & 18.794 & 18.735 & 18.508 & 18.485 & 18.396 & 18.330 & 18.332 & 18.221 & 18.036 & 18.078 & 18.071 & 18.168\\
   & 0.082 &  0.071 &  0.059 &  0.063 &  0.053 &  0.056 &  0.059 &  0.065 &  0.071 &  0.063 &  0.080 &  0.082 &  0.161 \\
\end{tabular}
\end{table}
}

\clearpage
{\small
\setcounter{table}{1}
\begin{table}[htb]
\caption{Continued}
\vspace {0.3cm}
\begin{tabular}{cccccccccccccc}
\hline
\hline
 No. & 03  &  04 &  05 &  06 &  07 &  08 &  09 &  10 &  11 &  12 &  13 &  14 &
15\\
(1)    & (2) & (3) & (4) & (5) & (6) & (7) & (8) & (9) & (10) & (11) & (12) & (13) & (14)\\
\hline
     113 & 19.868 & 19.545 & 19.527 & 19.685 & 19.362 & 19.618 & 19.706 & 19.577 & 19.689 & 19.498 & 19.968 & 20.229 & 20.099\\
   & 0.109 &  0.074 &  0.068 &  0.089 &  0.066 &  0.083 &  0.101 &  0.122 &  0.146 &  0.142 &  0.324 &  0.438 &  0.812 \\
     115 & 17.867 & 17.660 & 17.796 & 17.740 & 17.836 & 17.857 & 17.920 & 17.798 & 17.940 & 17.874 & 17.975 & 17.971 & 17.635\\
   & 0.023 &  0.018 &  0.018 &  0.020 &  0.019 &  0.023 &  0.027 &  0.029 &  0.038 &  0.039 &  0.055 &  0.063 &  0.078 \\
     117 & 19.060 & 18.904 & 18.837 & 18.710 & 18.726 & 18.790 & 18.718 & 18.741 & 18.765 & 18.636 & 18.809 & 18.355 & 18.748\\
   & 0.164 &  0.160 &  0.132 &  0.149 &  0.156 &  0.177 &  0.196 &  0.213 &  0.252 &  0.228 &  0.362 &  0.253 &  0.433 \\
     118 & 18.322 & 18.002 & 17.883 & 17.658 & 17.515 & 17.387 & 17.274 & 17.129 & 17.049 & 17.104 & 16.974 & 16.722 & 16.782\\
   & 0.163 &  0.154 &  0.123 &  0.126 &  0.091 &  0.094 &  0.087 &  0.084 &  0.084 &  0.084 &  0.089 &  0.066 &  0.089 \\
     119 & 18.481 & 18.162 & 18.088 & 17.866 & 17.719 & 17.599 & 17.540 & 17.336 & 17.262 & 17.306 & 17.163 & 16.898 & 16.947\\
   & 0.160 &  0.154 &  0.122 &  0.126 &  0.093 &  0.094 &  0.096 &  0.082 &  0.086 &  0.081 &  0.086 &  0.063 &  0.089 \\
     120 & 18.281 & 17.983 & 18.108 & 18.002 & 18.041 & 18.003 & 18.165 & 18.019 & 18.252 & 18.369 & 18.309 & 19.015 & 19.853\\
   & 0.261 &  0.236 &  0.234 &  0.280 &  0.234 &  0.276 &  0.309 &  0.308 &  0.427 &  0.429 &  0.470 &  0.904 &  2.539 \\
     121 & 18.602 & 18.290 & 18.297 & 18.184 & 18.466 & 18.422 & 17.962 & 18.808 & 19.161 & 19.213 & 19.548 & 19.123 & 20.396\\
   & 0.189 &  0.148 &  0.139 &  0.137 &  0.147 &  0.154 &  0.189 &  0.221 &  0.370 &  0.412 &  0.739 &  0.600 &  2.042 \\
     122 & 17.297 & 17.137 & 17.189 & 17.132 & 17.025 & 17.054 & 17.125 & 17.010 & 17.031 & 16.954 & 16.881 & 16.804 & 16.758\\
   & 0.165 &  0.145 &  0.129 &  0.144 &  0.096 &  0.123 &  0.200 &  0.155 &  0.186 &  0.134 &  0.146 &  0.135 &  0.191 \\
     124 & 18.820 & 18.849 & 19.003 & 18.906 & 19.068 & 19.208 & 19.177 & 19.057 & 18.998 & 19.262 & 19.231 & 19.251 & 18.488\\
   & 0.144 &  0.161 &  0.155 &  0.173 &  0.163 &  0.208 &  0.218 &  0.194 &  0.189 &  0.276 &  0.293 &  0.324 &  0.239 \\  
     125 & 18.412 & 18.348 & 18.463 & 18.429 & 18.485 & 18.500 & 18.520 & 18.491 & 18.812 & 18.746 & 18.457 & 18.364 & 18.855\\
   & 0.134 &  0.157 &  0.160 &  0.192 &  0.183 &  0.220 &  0.258 &  0.273 &  0.423 &  0.368 &  0.309 &  0.277 &  0.588 \\
     126 & 20.649 & 19.648 & 19.545 & 19.327 & 18.988 & 18.935 & 19.072 & 18.551 & 18.416 & 18.252 & 18.116 & 17.992 & 18.019\\
   & 0.876 &  0.393 &  0.362 &  0.366 &  0.223 &  0.235 &  0.312 &  0.190 &  0.188 &  0.151 &  0.153 &  0.146 &  0.181 \\
     127 & 15.712 & 15.650 & 15.796 & 15.767 & 15.971 & 15.950 & 15.990 & 15.966 & 15.994 & 16.108 & 16.167 & 16.250 & 16.078\\
   & 0.035 &  0.037 &  0.038 &  0.044 &  0.065 &  0.070 &  0.104 &  0.095 &  0.104 &  0.107 &  0.126 &  0.139 &  0.148 \\
     128 & 18.885 & 18.545 & 18.441 & 18.464 & 18.246 & 18.375 & 18.286 & 18.155 & 17.886 & 17.819 & 17.829 & 17.690 & 17.531\\
   & 0.429 &  0.352 &  0.268 &  0.367 &  0.263 &  0.342 &  0.327 &  0.330 &  0.298 &  0.268 &  0.291 &  0.281 &  0.296 \\
     129 & 18.234 & 18.107 & 18.174 & 18.193 & 17.865 & 18.039 & 18.037 & 17.910 & 17.675 & 17.642 & 17.357 & 17.286 & 17.284\\
   & 0.204 &  0.214 &  0.203 &  0.280 &  0.159 &  0.235 &  0.276 &  0.261 &  0.236 &  0.213 &  0.167 &  0.177 &  0.204 \\
     130 & 17.671 & 17.323 & 17.481 & 17.221 & 17.431 & 17.063 & 16.851 & 16.810 & 16.780 & 17.125 & 16.663 & 16.678 & 16.598\\
   & 0.089 &  0.078 &  0.079 &  0.089 &  0.087 &  0.080 &  0.066 &  0.075 &  0.086 &  0.114 &  0.087 &  0.093 &  0.108 \\
     131 & 18.270 & 17.870 & 17.868 & 17.746 & 17.539 & 17.479 & 17.316 & 17.315 & 17.320 & 17.336 & 17.205 & 17.238 & 17.084\\
   & 0.113 &  0.089 &  0.074 &  0.079 &  0.052 &  0.054 &  0.049 &  0.052 &  0.061 &  0.058 &  0.069 &  0.077 &  0.085 \\
     132 & 18.892 & 18.955 & 18.947 & 18.861 & 18.929 & 18.856 & 18.697 & 18.647 & 18.495 & 18.548 & 18.492 & 18.581 & 18.247\\
   & 0.211 &  0.216 &  0.187 &  0.203 &  0.170 &  0.174 &  0.176 &  0.182 &  0.172 &  0.180 &  0.227 &  0.256 &  0.256 \\
     133 & 18.650 & 18.490 & 18.531 & 18.497 & 18.402 & 18.395 & 18.242 & 18.170 & 18.113 & 18.064 & 17.988 & 17.981 & 17.860\\
   & 0.123 &  0.127 &  0.124 &  0.162 &  0.117 &  0.135 &  0.123 &  0.124 &  0.127 &  0.114 &  0.136 &  0.127 &  0.157 \\
\end{tabular}
\end{table}
}

\clearpage
{\small
\setcounter{table}{1}
\begin{table}[htb]
\caption{Continued}
\vspace {0.3cm}
\begin{tabular}{cccccccccccccc}
\hline
\hline
 No. & 03  &  04 &  05 &  06 &  07 &  08 &  09 &  10 &  11 &  12 &  13 &  14 &
15\\
(1)    & (2) & (3) & (4) & (5) & (6) & (7) & (8) & (9) & (10) & (11) & (12) & (13) & (14)\\
\hline 
     135 & 19.484 & 19.079 & 19.000 & 19.412 & 19.010 & 18.905 & 18.500 & 18.947 & 18.643 & 18.708 & 18.632 & 18.785 & 18.347\\
   & 0.355 &  0.290 &  0.226 &  0.425 &  0.236 &  0.220 &  0.153 &  0.258 &  0.188 &  0.248 &  0.261 &  0.356 &  0.294 \\
     136 & 19.374 & 19.079 & 19.099 & 18.960 & 18.877 & 18.905 & 18.869 & 18.865 & 18.835 & 18.719 & 18.731 & 18.949 & 18.561\\
   & 0.213 &  0.216 &  0.201 &  0.236 &  0.187 &  0.224 &  0.205 &  0.242 &  0.258 &  0.221 &  0.275 &  0.342 &  0.303 \\
     137 & 18.374 & 18.095 & 18.196 & 18.136 & 18.121 & 18.130 & 18.066 & 18.137 & 18.138 & 18.090 & 18.139 & 17.906 & 18.133\\
   & 0.072 &  0.065 &  0.067 &  0.074 &  0.064 &  0.075 &  0.082 &  0.090 &  0.103 &  0.101 &  0.140 &  0.106 &  0.188 \\
     139 & 18.633 & 18.520 & 18.578 & 18.535 & 18.458 & 18.413 & 18.351 & 18.317 & 18.227 & 18.332 & 18.352 & 17.962 & 18.158\\
   & 0.069 &  0.062 &  0.063 &  0.066 &  0.056 &  0.061 &  0.067 &  0.069 &  0.072 &  0.082 &  0.116 &  0.079 &  0.149 \\
     141 & 16.069 & 15.987 & 16.128 & 16.155 & 16.240 & 16.306 & 16.371 & 16.357 & 16.419 & 16.395 & 16.284 & 16.346 & 16.396\\
   & 0.033 &  0.043 &  0.041 &  0.051 &  0.041 &  0.052 &  0.078 &  0.064 &  0.075 &  0.067 &  0.071 &  0.071 &  0.094 \\
     142 & 15.743 & 15.699 & 15.809 & 15.800 & 15.863 & 15.856 & 15.761 & 15.617 & 15.451 & 15.305 & 15.310 & 15.184 & 14.932\\
   & 0.012 &  0.011 &  0.010 &  0.014 &  0.010 &  0.011 &  0.012 &  0.010 &  0.009 &  0.007 &  0.009 &  0.008 &  0.009 \\
     144 & 19.991 & 19.810 & 19.717 & 19.464 & 19.386 & 19.214 & 18.917 & 19.089 & 19.036 & 19.060 & 19.145 & 19.179 & 19.271\\
   & 0.280 &  0.216 &  0.187 &  0.175 &  0.124 &  0.123 &  0.125 &  0.143 &  0.130 &  0.126 &  0.239 &  0.206 &  0.391 \\
     145 & 20.109 & 19.798 & 19.680 & 19.592 & 19.385 & 19.249 & 19.422 & 19.360 & 19.483 & 19.586 & 19.931 & 19.743 & 21.150\\
   & 0.255 &  0.202 &  0.169 &  0.188 &  0.136 &  0.141 &  0.179 &  0.192 &  0.238 &  0.285 &  0.452 &  0.431 &  2.438 \\ 
     146 & 18.574 & 18.249 & 18.294 & 18.135 & 18.350 & 18.303 & 18.350 & 18.470 & 18.363 & 18.436 & 18.446 & 18.493 & 18.114\\
   & 0.278 &  0.245 &  0.235 &  0.272 &  0.246 &  0.273 &  0.393 &  0.357 &  0.387 &  0.358 &  0.432 &  0.426 &  0.328 \\
     147 & 18.439 & 18.360 & 18.441 & 18.356 & 18.420 & 18.419 & 18.328 & 18.393 & 18.521 & 18.541 & 18.809 & 18.709 & 18.733\\
   & 0.033 &  0.031 &  0.027 &  0.027 &  0.026 &  0.033 &  0.031 &  0.042 &  0.046 &  0.060 &  0.134 &  0.100 &  0.181 \\
     148 & 18.016 & 17.901 & 17.520 & 18.051 & 18.136 & 18.184 & 17.009 & 18.122 & 18.111 & 18.200 & 18.618 & 17.935 & 18.093\\
   & 0.076 &  0.072 &  0.041 &  0.068 &  0.072 &  0.081 &  0.040 &  0.108 &  0.119 &  0.130 &  0.238 &  0.134 &  0.207 \\
     150 & 17.541 & 17.380 & 17.432 & 17.361 & 17.408 & 17.432 & 17.426 & 17.410 & 17.444 & 17.470 & 17.575 & 17.479 & 17.491\\
   & 0.064 &  0.062 &  0.053 &  0.070 &  0.067 &  0.079 &  0.085 &  0.093 &  0.104 &  0.101 &  0.128 &  0.119 &  0.147 \\
     151 & 17.717 & 17.493 & 17.502 & 17.390 & 17.333 & 17.305 & 17.257 & 17.200 & 17.152 & 17.097 & 17.067 & 16.968 & 16.941\\
   & 0.063 &  0.063 &  0.055 &  0.067 &  0.054 &  0.061 &  0.062 &  0.062 &  0.064 &  0.059 &  0.068 &  0.063 &  0.080 \\
     152 & 19.190 & 18.928 & 18.904 & 18.942 & 18.840 & 18.901 & 18.746 & 18.870 & 18.701 & 18.730 & 18.806 & 18.683 & 18.534\\
   & 0.137 &  0.130 &  0.128 &  0.156 &  0.125 &  0.148 &  0.145 &  0.177 &  0.163 &  0.171 &  0.249 &  0.206 &  0.264 \\
     153 & 18.949 & 18.733 & 18.681 & 18.616 & 18.538 & 18.549 & 18.387 & 18.491 & 18.573 & 18.362 & 18.095 & 18.405 & 18.038\\
   & 0.132 &  0.129 &  0.125 &  0.116 &  0.105 &  0.104 &  0.141 &  0.113 &  0.139 &  0.126 &  0.127 &  0.175 &  0.162 \\
     154 & 17.890 & 17.734 & 17.789 & 17.698 & 17.718 & 17.713 & 17.575 & 17.611 & 17.569 & 17.512 & 17.407 & 17.338 & 17.323\\
   & 0.045 &  0.041 &  0.040 &  0.045 &  0.040 &  0.047 &  0.066 &  0.055 &  0.067 &  0.062 &  0.067 &  0.062 &  0.107 \\
     155 & 17.677 & 17.509 & 17.548 & 17.530 & 17.478 & 17.503 & 17.501 & 17.539 & 17.504 & 17.467 & 17.533 & 17.428 & 17.529\\
   & 0.030 &  0.028 &  0.028 &  0.036 &  0.032 &  0.036 &  0.041 &  0.049 &  0.058 &  0.057 &  0.083 &  0.076 &  0.103 \\
     156 & 18.492 & 18.356 & 18.375 & 18.248 & 18.267 & 18.233 & 18.327 & 18.297 & 18.387 & 18.350 & 18.244 & 18.218 & 18.651\\
   & 0.101 &  0.111 &  0.102 &  0.116 &  0.104 &  0.109 &  0.132 &  0.135 &  0.153 &  0.149 &  0.171 &  0.164 &  0.282 \\ 
\end{tabular}
\end{table}
}

\clearpage
{\small
\setcounter{table}{1}
\begin{table}[htb]
\caption{Continued}
\vspace {0.3cm}
\begin{tabular}{cccccccccccccc}
\hline
\hline
 No. & 03  &  04 &  05 &  06 &  07 &  08 &  09 &  10 &  11 &  12 &  13 &  14 &
15\\
(1)    & (2) & (3) & (4) & (5) & (6) & (7) & (8) & (9) & (10) & (11) & (12) & (13) & (14)\\
\hline
     157 & 20.023 & 19.557 & 19.483 & 19.239 & 19.180 & 19.180 & 19.181 & 19.103 & 19.149 & 19.052 & 19.202 & 19.499 & 18.769\\
   & 0.401 &  0.310 &  0.247 &  0.264 &  0.210 &  0.229 &  0.258 &  0.247 &  0.303 &  0.266 &  0.370 &  0.532 &  0.301 \\
     158 & 15.950 & 15.913 & 15.947 & 16.070 & 16.158 & 16.146 & 15.869 & 16.074 & 15.849 & 15.869 & 15.813 & 15.508 & 15.466\\
   & 0.044 &  0.050 &  0.045 &  0.065 &  0.051 &  0.058 &  0.057 &  0.059 &  0.055 &  0.048 &  0.054 &  0.043 &  0.046 \\
     159 & 16.699 & 16.684 & 16.909 & 16.777 & 17.034 & 16.943 & 16.937 & 16.984 & 16.903 & 17.134 & 16.907 & 16.910 & 16.877\\
   & 0.034 &  0.032 &  0.035 &  0.036 &  0.035 &  0.036 &  0.056 &  0.049 &  0.052 &  0.056 &  0.063 &  0.064 &  0.083 \\
     160 & 18.762 & 18.558 & 18.638 & 18.565 & 18.521 & 18.452 & 18.495 & 18.332 & 18.366 & 18.334 & 18.042 & 18.183 & 18.043\\
   & 0.098 &  0.088 &  0.082 &  0.091 &  0.072 &  0.079 &  0.098 &  0.082 &  0.090 &  0.091 &  0.110 &  0.119 &  0.135 \\
     161 & 19.364 & 19.028 & 18.869 & 18.660 & 18.457 & 18.339 & 18.180 & 18.074 & 17.956 & 17.942 & 17.951 & 17.631 & 17.631\\
   & 0.077 &  0.054 &  0.039 &  0.034 &  0.033 &  0.033 &  0.034 &  0.036 &  0.035 &  0.038 &  0.059 &  0.043 &  0.078 \\
     162 & 20.196 & 20.274 & 20.054 & 19.939 & 19.828 & 19.798 & 19.283 & 19.342 & 19.375 & 19.258 & 19.030 & 19.141 & 18.773\\
   & 0.130 &  0.106 &  0.095 &  0.089 &  0.078 &  0.087 &  0.097 &  0.080 &  0.089 &  0.087 &  0.143 &  0.135 &  0.168 \\ 
\hline
\end{tabular}
\end{table}
}

\clearpage
\setcounter{table}{2}
\begin{table}[htb]
\caption{Comparison of Cluster Photometry with Previous Measurements}
\vspace {0.3cm}
\begin{tabular}{ccc|ccc}
\hline
\hline
No. & $V$ (Chandar et al.)  & $V$ (BATC) & 	No. & $V$ (Chandar et al.)  & $V$ (BATC) \\
(1)    & (2) & (3) & 	(1)    & (2) & (3) \\
\hline
     62...... & 18.769 $\pm$  0.005 & 19.332 $\pm$  0.240 &   108...... & 19.162 $\pm$  0.006  &19.262 $\pm$  0.216 \\
     64...... & 19.007 $\pm$  0.008 & 19.107 $\pm$  0.147 &   109...... & 18.527 $\pm$  0.000  &17.802 $\pm$  0.079 \\
     67...... & 17.446 $\pm$  0.002 & 17.515 $\pm$  0.056 &   110...... & 18.515 $\pm$  0.003  &18.596 $\pm$  0.078 \\
     68...... & 17.963 $\pm$  0.003 & 17.953 $\pm$  0.082 &   112...... & 18.625 $\pm$  0.004  &18.580 $\pm$  0.091 \\
     69...... & 18.541 $\pm$  0.005 & 18.697 $\pm$  0.179 &   113...... & 19.269 $\pm$  0.006  &19.443 $\pm$  0.122 \\
     71...... & 18.822 $\pm$  0.004 & 19.134 $\pm$  0.183 &   115...... & 19.648 $\pm$  0.007  &17.857 $\pm$  0.033 \\
     72...... & 18.321 $\pm$  0.003 & 18.293 $\pm$  0.089 &   117...... & 18.363 $\pm$  0.006  &18.759 $\pm$  0.261 \\
     73...... & 19.430 $\pm$  0.007 & 19.536 $\pm$  0.183 &   118...... & 17.945 $\pm$  0.005  &17.662 $\pm$  0.162 \\
     74...... & 18.780 $\pm$  0.003 & 18.827 $\pm$  0.119 &   119...... & 18.247 $\pm$  0.006  &17.864 $\pm$  0.164 \\
     75...... & 19.534 $\pm$  0.007 & 19.105 $\pm$  0.256 &   120...... & 18.169 $\pm$  0.000  &18.100 $\pm$  0.414 \\
     76...... & 19.687 $\pm$  0.000 & 18.961 $\pm$  0.196 &   121...... & 18.431 $\pm$  0.010  &18.448 $\pm$  0.241 \\
     77...... & 18.778 $\pm$  0.003 & 18.153 $\pm$  0.059 &   122...... & 17.343 $\pm$  0.004  &17.109 $\pm$  0.182 \\
     78...... & 18.238 $\pm$  0.003 & 18.221 $\pm$  0.125 &   124...... & 18.859 $\pm$  0.009  &19.029 $\pm$  0.286 \\
     79...... & 18.969 $\pm$  0.005 & 18.945 $\pm$  0.210 &   125...... & 17.983 $\pm$  0.005  &18.521 $\pm$  0.316 \\
     83...... & 19.426 $\pm$  0.006 & 19.593 $\pm$  0.159 &   126...... & 18.518 $\pm$  0.007  &19.174 $\pm$  0.417 \\
     84...... & 19.705 $\pm$  0.006 & 20.023 $\pm$  0.247 &   127...... & 16.394 $\pm$  0.003  &15.971 $\pm$  0.102 \\
     86...... & 18.945 $\pm$  0.004 & 19.158 $\pm$  0.097 &   128...... & 17.841 $\pm$  0.010  &18.334 $\pm$  0.492 \\
     87...... & 19.041 $\pm$  0.006 & 19.037 $\pm$  0.080 &   129...... & 17.383 $\pm$  0.006  &17.974 $\pm$  0.325 \\
     88...... & 18.198 $\pm$  0.003 & 17.884 $\pm$  0.248 &   130...... & 17.838 $\pm$  0.006  &17.541 $\pm$  0.142 \\
     89...... & 18.538 $\pm$  0.004 & 18.393 $\pm$  0.269 &   131...... & 18.262 $\pm$  0.007  &17.684 $\pm$  0.095 \\
     91...... & 17.886 $\pm$  0.003 & 17.861 $\pm$  0.075 &   132...... & 18.678 $\pm$  0.014  &18.990 $\pm$  0.292 \\
     92...... & 18.605 $\pm$  0.008 & 18.683 $\pm$  0.171 &   133...... & 18.106 $\pm$  0.006  &18.494 $\pm$  0.213 \\
     93...... & 19.105 $\pm$  0.014 & 19.449 $\pm$  0.350 &   135...... & 18.826 $\pm$  0.013  &19.233 $\pm$  0.445 \\
     94...... & 18.478 $\pm$  0.005 & 18.516 $\pm$  0.109 &   136...... & 18.807 $\pm$  0.011 & 18.954 $\pm$  0.336 \\
     95...... & 18.289 $\pm$  0.004 & 17.872 $\pm$  0.065 &   137...... & 18.011 $\pm$  0.006 & 18.182 $\pm$  0.112 \\
     96...... & 19.075 $\pm$  0.009 & 19.486 $\pm$  0.312 &   139...... & 18.223 $\pm$  0.004 & 18.556 $\pm$  0.097 \\
     97...... & 18.283 $\pm$  0.006 & 18.455 $\pm$  0.117 &   141...... & 16.281 $\pm$  0.002 & 16.250 $\pm$  0.074 \\
     99...... & 18.154 $\pm$  0.006  &18.443 $\pm$  0.262 &   142...... & 15.854 $\pm$  0.001 & 15.904 $\pm$  0.018 \\
    100...... & 17.697 $\pm$  0.007  &17.183 $\pm$  0.085 &   144...... & 19.055 $\pm$  0.014 & 19.526 $\pm$  0.220 \\
    101...... & 18.721 $\pm$  0.012  &19.097 $\pm$  0.498 &   145...... & 19.329 $\pm$  0.016 & 19.555 $\pm$  0.242 \\
    103...... & 18.525 $\pm$  0.011  &18.874 $\pm$  0.374 &   146...... & 18.577 $\pm$  0.012 & 18.355 $\pm$  0.422 \\
    107...... & 18.378 $\pm$  0.003  &18.459 $\pm$  0.079 &   147...... & 18.423 $\pm$  0.007 & 18.459 $\pm$  0.045 \\ 
\end{tabular}
\end{table}

\clearpage
\setcounter{table}{2}
\begin{table}[htb]
\caption{Continued}
\vspace {0.3cm}
\begin{tabular}{ccc|ccc}
\hline
\hline
No. & $V$ (Chandar et al.)  & $V$ (BATC) & No. & $V$ (Chandar et al.)  & $V$ (BATC)\\
(1)    & (2) & (3) & (1) & (2) & (3)\\
\hline
    148...... & 17.714 $\pm$  0.006 & 18.152 $\pm$  0.120 & 156...... & 18.341 $\pm$  0.008 & 18.331 $\pm$  0.177 \\
    150...... & 17.398 $\pm$  0.004 & 17.444 $\pm$  0.115 & 157...... & 18.979 $\pm$  0.012 & 19.258 $\pm$  0.369 \\
    151...... & 17.242 $\pm$  0.004 & 17.419 $\pm$  0.095 & 158...... & 16.191 $\pm$  0.002 & 16.192 $\pm$  0.091 \\
    152...... & 18.632 $\pm$  0.009 & 18.912 $\pm$  0.223 & 159...... & 17.000 $\pm$  0.003 & 17.039 $\pm$  0.058 \\
    153...... & 18.610 $\pm$  0.009 & 18.619 $\pm$  0.176 & 160...... & 18.458 $\pm$  0.007 & 18.617 $\pm$  0.127 \\
    154...... & 17.924 $\pm$  0.005 & 17.772 $\pm$  0.070 & 161...... & 18.749 $\pm$  0.005 & 18.620 $\pm$  0.055 \\ 
    155...... & 17.504 $\pm$  0.003 & 17.546 $\pm$  0.055 & 162...... & 19.920 $\pm$  0.014 & 19.933 $\pm$  0.135 \\
\hline
\end{tabular}
\end{table}

\clearpage
\setcounter{table}{3}
\begin{table}[htb]
\caption[]{Age Distribution of 78 Star Clusters}
\vspace {0.3cm}
\begin{tabular}{ccc|ccc}
\hline
\hline
 No. & Metallicity ($Z$)& Age ([$\log$ yr]) & No. & Metallicity ($Z$)& Age ([$\log$ yr])\\
 (1) & (2) & (3) & (1) & (2) & (3) \\
\hline
     62...... & 0.00040 &  9.279 &          108...... & 0.00400 &  8.507\\
     64...... & 0.00400 &  8.806 &          109...... & 0.00040 &  8.657\\
     67...... & 0.00400 &  7.720 &          110...... & 0.00040 &  8.957\\
     68...... & 0.00400 &  7.021 &          112...... & 0.00040 &  9.207\\
     69...... & 0.00040 &  9.760 &          113...... & 0.00040 &  7.806\\
     71...... & 0.02000 &  8.606 &          115...... & 0.02000 &  6.840\\
     72...... & 0.02000 &  9.107 &          117...... & 0.00400 &  8.009\\
     73...... & 0.02000 &  9.107 &          118...... & 0.02000 &  9.155\\
     74...... & 0.00400 &  9.322 &          119...... & 0.02000 &  9.155\\
     75...... & 0.00040 &  8.757 &          120...... & 0.00400 &  6.600\\
     76...... & 0.00400 &  8.757 &          121...... & 0.02000 &  6.620\\
     77...... & 0.02000 &  9.009 &          122...... & 0.00400 &  8.307\\
     78...... & 0.02000 &  6.800 &          124...... & 0.02000 &  6.860\\
     79...... & 0.02000 &  9.107 &          125...... & 0.00400 &  6.860\\
     83...... & 0.02000 &  6.940 &          126...... & 0.02000 &  9.954\\
     84...... & 0.02000 &  6.860 &          127...... & 0.00040 &  7.220\\
     86...... & 0.00040 &  9.155 &          128...... & 0.00400 &  9.107\\
     87...... & 0.00040 & 10.061 &          129...... & 0.02000 &  6.940\\
     88...... & 0.00400 &  6.480 &          130...... & 0.00400 &  9.057\\
     89...... & 0.00040 &  6.580 &          131...... & 0.00040 &  9.301\\
     91...... & 0.00040 &  9.255 &          132...... & 0.02000 &  6.980\\
     92...... & 0.00400 &  7.699 &          133...... & 0.02000 &  6.960\\
     93...... & 0.00400 &  6.600 &          135...... & 0.02000 &  6.940\\
     94...... & 0.00040 &  8.356 &          136...... & 0.00040 &  8.806\\
     95...... & 0.02000 &  6.940 &          137...... & 0.02000 &  8.057\\
     96...... & 0.02000 &  6.920 &          139...... & 0.02000 &  7.179\\
     97...... & 0.00400 & 10.279 &          141...... & 0.00040 &  6.660\\
     99...... & 0.02000 &  9.107 &          142...... & 0.02000 &  6.940\\
    100...... & 0.02000 &  6.680 &          144...... & 0.00040 &  9.107\\
    101...... & 0.02000 &  8.057 &          145...... & 0.00040 &  8.009\\
    103...... & 0.00400 &  6.620 &          146...... & 0.00040 &  8.009\\
    107...... & 0.00400 &  8.857 &          147...... & 0.02000 &  6.760\\ 
\end{tabular}
\end{table}

\clearpage
\setcounter{table}{3}
\begin{table}[htb]
\caption[]{Continued}
\vspace {0.3cm}
\begin{tabular}{ccc|ccc}
\hline
\hline
 No. & Metallicity ($Z$)& Age ([$\log$ yr]) & No. & Metallicity ($Z$)& Age ([$\log$ yr])\\
 (1) & (2) & (3) & (1) & (2) & (3) \\
\hline
    148...... & 0.02000 &  6.840 &          156...... & 0.00040 &  8.009\\
    150...... & 0.00040 &  8.009 &          157...... & 0.00040 &  8.957\\
    151...... & 0.00400 &  8.757 &          158...... & 0.02000 &  6.980\\
    152...... & 0.00400 &  8.356 &          159...... & 0.00400 &  7.179\\
    153...... & 0.00040 &  8.906 &          160...... & 0.00040 &  8.507\\
    154...... & 0.00040 &  8.507 &          161...... & 0.02000 &  9.225\\
    155...... & 0.00400 &  6.960 &          162...... & 0.00400 & 10.283\\
\hline
\end{tabular}
\end{table}

\end{document}